\documentclass[12pt]{article}
\usepackage{epsfig}
\begin{document}
\newcommand{\noi}{\noindent}
\newtheorem{opgave}{Excercise}
\newcommand{\bop}[1]{\begin{opgave}\begin{rm}{\bf#1} \newline\noi}
\newcommand{\eop}{\end{rm}\end{opgave}}

\newcommand{\nc}{\newcommand}
\nc{\dia}[1]{\mbox{{\Large{\bf diag}}}[#1]}
\nc{\diag}[3]{\raisebox{#3cm}{\epsfig{figure=#1.eps,width=#2cm}}}
\nc{\bq}{\begin{equation}}
\nc{\eq}{\end{equation}}
\nc{\bqa}{\begin{eqnarray}} 
\nc{\eqa}{\end{eqnarray}}
\nc{\eqan}{\nonumber\end{eqnarray}}
\nc{\nl}{\nonumber \\}
\nc{\grf}{Green's function}
\nc{\grfs}{Green's functions}
\nc{\cgrf}{connected Green's function}
\nc{\cgrfs}{connected Green's functions}
\nc{\f}{\varphi}
\nc{\exb}[1]{\exp\!\left(#1\right)}
\nc{\avg}[1]{\left\langle #1\right\rangle}
\nc{\suml}{\sum\limits}
\nc{\prodl}{\prod\limits}
\nc{\intl}{\int\limits}
\nc{\ddv}[1]{{\partial\over\partial #1}}
\nc{\ddvv}[2]{{\partial^{#1}\over\left(\partial #2\right)^{#1}}}
\nc{\la}{\lambda}
\nc{\La}{\Lambda}
\nc{\eqn}[1]{Eq.(\ref{#1})}
\nc{\eqns}[1]{Eqs.(\ref{#1})}
\nc{\appndix}[3]{\section{{#2}\label{#3}}\input{#3}}
\nc{\cala}{{\cal A}}
\nc{\calk}{{\cal K}}
\nc{\calb}{{\cal B}}
\nc{\cale}{{\cal E}}
\nc{\calf}{{\cal F}}
\nc{\calh}{{\cal H}}
\nc{\calc}{{\cal C}}
\nc{\cald}{{\cal D}}
\nc{\calp}{{\cal P}}
\nc{\calphi}{\Psi}
\nc{\lra}{\;\leftrightarrow\;}
\nc{\stapel}[1]{\begin{tabular}{c} #1\end{tabular}}
\nc{\al}{\alpha}
\nc{\be}{\beta}
\nc{\g}{\gamma}
\nc{\ga}{\gamma}
\nc{\ka}{\kappa}
\nc{\om}{\omega}
\nc{\si}{\sigma}
\nc{\Si}{\Sigma}
\nc{\ro}{\rho}
\nc{\vph}{\vphantom{{A^A\over A_A}}}
\nc{\wph}{\vphantom{{A^A}}}
\nc{\order}[1]{{\cal O}\left(#1\right)}
\nc{\De}{\Delta}
\nc{\de}{\delta}
\nc{\vx}{\vec{x}}
\nc{\vxi}{\vec{\xi}}
\nc{\vy}{\vec{y}}
\nc{\vk}{\vec{k}}
\nc{\vn}{\vec{n}}
\nc{\vq}{\vec{q}}
\nc{\vp}{\vec{p}}
\nc{\vz}{\vec{z}}
\nc{\vt}{\vec{t}}
\nc{\nv}{\vec{n}}
\nc{\ve}{\vec{e}}
\nc{\lijst}[1]{\begin{center}
  \fbox{\begin{minipage}{12cm}{#1}\end{minipage}}\end{center}}
\nc{\lijstx}[2]{\begin{center}
  \fbox{\begin{minipage}{#1cm}{#2}\end{minipage}}\end{center}}
\nc{\eps}{\epsilon}
\nc{\caput}[2]{\chapter{#1}\input{#2}}
\nc{\pd}{\partial}
\nc{\dtil}{\tilde{d}}
\nc{\vskp}{\vspace*{\baselineskip}}
\nc{\coos}{co\-ord\-in\-at\-es}
\nc{\calm}{{\cal M}}
\nc{\calj}{{\cal J}}
\nc{\call}{{\cal L}}
\nc{\calt}{{\cal T}}
\nc{\calu}{{\cal U}}
\nc{\calw}{{\cal W}}
\nc{\msq}{\langle|\calm|^2\rangle}
\nc{\nsym}{F_{\mbox{{\tiny{symm}}}}}
\nc{\dm}[1]{\mbox{\bf dim}\!\left[#1\right]}
\nc{\fourv}[4]{\left(\begin{tabular}{c}
  $#1$\\$#2$\\$#3$\\$#4$\end{tabular}\right)}
\nc{\driev}[3]{\left(\begin{tabular}{c}
  $#1$\\$#2$\\$#3$\end{tabular}\right)}
\nc{\fs}[1]{/\!\!\!#1}
\nc{\dbar}[1]{{\overline{#1}}}
\nc{\tr}[1]{\mbox{Tr}\left(#1\right)}
\nc{\row}[4]{$#1$ & $#2$ & $#3$ & $#4$}
\nc{\matv}[4]{\left(\begin{tabular}{cccc}
   #1 \\ #2 \\ #3 \\ #4\end{tabular}\right)}
\nc{\twov}[2]{\left(\begin{tabular}{cccc}
   $#1$ \\ $#2$\end{tabular}\right)}
\nc{\ubar}{\dbar{u}}
\nc{\vbar}{\dbar{v}}
\nc{\lng}{longitudinal}
\nc{\pol}{polarisation}
\nc{\longpol}{\lng\ \pol}
\nc{\bnum}{\begin{enumerate}}
\nc{\enum}{\end{enumerate}}
\nc{\nubar}{\overline{\nu}}
\nc{\mau}{{m_{\mbox{{\tiny U}}}}}
\nc{\mad}{{m_{\mbox{{\tiny D}}}}}
\nc{\qu}{{Q_{\mbox{{\tiny U}}}}}
\nc{\qd}{{Q_{\mbox{{\tiny D}}}}}
\nc{\vau}{{v_{\mbox{{\tiny U}}}}}
\nc{\aau}{{a_{\mbox{{\tiny U}}}}}
\nc{\aad}{{a_{\mbox{{\tiny D}}}}}
\nc{\vad}{{v_{\mbox{{\tiny D}}}}}
\nc{\mw}{{m_{\mbox{{\tiny W}}}}}
\nc{\mv}{{m_{\mbox{{\tiny V}}}}}
\nc{\mz}{{m_{\mbox{{\tiny Z}}}}}
\nc{\gz}{{\Gamma_{\mbox{{\tiny Z}}}}}
\nc{\mh}{{m_{\mbox{{\tiny H}}}}}
\nc{\gw}{{g_{\mbox{{\tiny W}}}}}
\nc{\gs}{{g_{\mbox{{\small s}}}}}
\nc{\qw}{{Q_{\mbox{{\tiny W}}}}}
\nc{\qe}{{Q_{\mbox{{\tiny e}}}}}
\nc{\qc}{{Q_{\mbox{{\tiny c}}}}}
\nc{\law}{{\Lambda_{\mbox{{\tiny W}}}}}
\nc{\obar}{\overline}
\nc{\guuh}{{g_{\mbox{{\tiny UUH}}}}}
\nc{\gtth}{{g_{\mbox{{\tiny ttH}}}}}
\nc{\gwwz}{{g_{\mbox{{\tiny WWZ}}}}}
\nc{\gwwh}{{g_{\mbox{{\tiny WWH}}}}}
\nc{\gwwhh}{{g_{\mbox{{\tiny WWHH}}}}}
\nc{\gzzh}{{g_{\mbox{{\tiny ZZH}}}}}
\nc{\gvvh}{{g_{\mbox{{\tiny VVH}}}}}
\nc{\ghhh}{{g_{\mbox{{\tiny HHH}}}}}
\nc{\ghhhh}{{g_{\mbox{{\tiny HHHH}}}}}
\nc{\gzzhh}{{g_{\mbox{{\tiny ZZHH}}}}}
\nc{\thw}{\theta_W}
\nc{\sw}{{s_{\mbox{{\tiny W}}}}}
\nc{\cw}{{c_{\mbox{{\tiny W}}}}}
\nc{\ward}[2]{\left.#1\right\rfloor_{#2}}
\nc{\vpa}{(1+\g^5)}
\nc{\vma}{(1-\g^5)}
\nc{\gwzh}{{g_{\mbox{{\tiny WZH}}}}}
\nc{\gudh}{{g_{\mbox{{\tiny UDH}}}}}
\nc{\gwcj}{{g_{\mbox{{\tiny Wcj}}}}}
\nc{\gwcjg}{{g_{\mbox{{\tiny Wcj$\g$}}}}}
\nc{\gwwcc}{{g_{\mbox{{\tiny WWcc}}}}}
\nc{\gzcc}{{g_{\mbox{{\tiny Zcc}}}}}
\nc{\none}[1]{ }
\nc{\Vir}{{V_{\mbox{{\small IR}}}}}
\nc{\kt}[1]{\left|#1\right\rangle}
\nc{\br}[1]{\left\langle#1\right|}
\nc{\bak}[3]{\left\langle #1 \left| #2 \right| #3\right\rangle}
\nc{\tw}{\tilde{w}}
\nc{\hDe}{\nabla}
\nc{\Omb}{\overline{\Omega}}
\nc{\qbar}{\bar{q}}
\nc{\leeg}[1]{}
\nc{\dilog}{\mbox{Li}_2}
\nc{\marx}[1]{\marginpar{\fbox{{\bf E\,\ref{#1}}}}}
\nc{\ddx}{\ddv{x}}
\nc{\ddy}{\ddv{y}}
\nc{\ktil}{\tilde{k}}
\nc{\etil}{\tilde{\eps}}

\begin{center}
{\bf {\Large Amplitudes, recursion relations and unitarity 
in the Abelian Higgs Model}}\\
\vspace*{\baselineskip} 
Ronald Kleiss\footnote{{\tt R.Kleiss@science.ru.nl}}
 and Oscar Boher Luna\footnote{{\tt oscar$\underline{\hphantom a}$bl92@hotmail.com}}\\
\vspace*{\baselineskip} 
Institute for Mathematics, Astrophysics and Particle Physics, Radboud University
Nijmegen, Heyendaalseweg 135, Nijmegen, the Netherlands\\
\vspace*{3\baselineskip} 
{\bf Abstract}\\ 
\vspace*{\baselineskip} 
\begin{minipage}{12cm}
The Abelian Higgs model forms an essential part of the electroweak standard
model: it is the sector containing only Z${}^0$ and Higgs bosons. We present a
diagram-based
proof of the tree-level unitarity of this model inside the {\em unitary gauge}, where
only physical degrees of freedom occur. We derive combinatorial
recursion relations for off-shell amplitudes in the massless approximation, 
which allows us to prove the cancellation of the first two orders in energy of
unitarity-violating high-energy behaviour for any tree-level amplitude in this model.
We describe a deformation of the amplitudes by extending the physical 
phase space to at least 7 spacetime dimensions, which leads to on-shell
recursion relations   \`a la BCFW. These lead to a simple proof that
all on-shell tree amplitudes obey partial-wave unitarity.  

\end{minipage}
\end{center}
\setcounter{footnote}{0}
\newpage
\section{Introduction} 
The Minimal Standard Model includes a few sectors that are
consistent theories in their own right. The best-known of these are QED and
QCD, but in addition we have the  ZH sector: it contains only the Z and the H
bosons, with their interactions. This is, in fact, nothing but the
Abelian Higgs model \cite{Abelianhiggsmodel}, the simplest example of a 
spontaneously broken gauge symmetry in quantum field theory.
The internal consistency of such models is of course uncontroversial \cite{tHV}. 
In particular they respect  unitarity, with which we 
mean the behaviour of on-shell scattering amplitudes
with energy when all masses and scattering
angles are kept constant: it implies unitarity in each partial wave separately.
This is not trivial, especially because of the longitudinal degrees of freedom
present in massive spin-1 particles. The best-known proof of perturbative
unitarity of the Abelian Higgs model is presented \cite{brs}. 
In that paper, the authors point out that unphysical fields must be involved in the case of a general gauge, and the main issue is to get rid of these fields, and of the gauge dependence, in
the finally resulting S-matrix elements. Other proofs, like that of the
equivalence theorem \cite{HelV} also typically rely on the Feynman-'t Hooft gauge. 
There is, however, another way to view the Abelian Higgs model. Rather than starting with
the unbroken theory, which is a gauge theory, we may as well simply regard the
broken Lagrangian `as given', that is a theory containing two massive particles, with spins 0 and 1,
without worrying where it came from. Massive spin-1 theories do not suffer from the
necessity of fixing a gauge, since there is no gauge symmetry. It ought therefore to be possible
to prove unitarity of the amplitudes directly using only the physical fields, with the
Proca propagator\footnote{In gauge theory language, the unitary gauge.}, at least at the tree
level. In higher loops (which we do not consider here), the effect of the Faddeev-Popov ghosts
can be implemented by introducing counterterms proportional to the space-time volume,
as described {\it eg.\/} in \cite{knetter}\footnote{These counterterms arise from the
infinite-momentum limit of 1PI diagrams where a closed Z loop couples to an arbitrary number
of H legs, thus giving rise to a non-polynomial Higgs counterterm Lagrangian.}. This is the
approach we adopt here: we shall use only the physical Z and H fields, and the unitary-gauge
propagator for the Z.

The study of multi-leg amplitudes is a flourishing field. Excellent didactic reviews
are for instance \cite{elvangreview,weinzierl}. These mainly discuss theories with a
high degree of symmetry (with $N=4$ super-Yang-Mills as an extreme example), whereas
we are dealing with a theory with very little symmetry, and with explicitly massive
particles. In addition, the approach of choice is
to express all fields in terms of massless (Weyl) spinors and employ the arsenal of techniques
available for such formulations. In the spirit of the previous paragraph, we hold that it ought
to be possible to restrict ourselves to (scalar and) vector fields only: no spinors will
intrude in our derivations.\\

Partial-wave unitarity requires cross sections
at some energy scale $E$ to decrease as $E^{-2}$ when $E$
becomes large, and all angles are kept fixed \cite{zuber,peskin}. Since for
an $n$-point amplitude $\calm_n$, relevant to $2\to(n-2)$ processes, the 
concomitant phase space has dimensionality\footnote{Each of the $n-2$ final-state momenta
contributes $E^2$, and the delta function imposing four-momentum conservation scales as
$E^{-4}$.}
 $E^{2n-8}$, acceptable unitarity
(high-energy) behaviour implies 
\bq
\calm_n \sim E^{4-n}
\eq
at high energies. As we shall show, power counting gives us a behaviour
up to $E^{+2}$ for amplitudes in the ZH sector, so cancellations over
many orders of magnitude (powers of $E$ over some mass) must occur for large-$n$ 
amplitudes. At the tree level, an amplitude like 
that for ZZ$\;\to\;$4Z+6H is based
on  649,271,700 diagrams and calls for a cancellation over 10 orders of magnitude: 
clearly we must be as systematic as possible. 

Our strategy in this paper
will be as follows. We shall first establish effective Feynman rules that
describe {\em off}-shell amplitudes at the $E^2$ level, that is, the most dangerous
behaviour with energy. The Schwinger-Dyson equations (SDe) of the model
provide recursion relations between these amplitudes which have surprisingly
simple solutions. The vanishing of the $E^2$ terms is then immediately obvious.
The $E^0$ terms can be obtained from these off-shell amplitudes by including
the effects of nonzero masses in a perturbative approach, and we shall show 
that in first order these vanish as well, provided that the Higgs self-interactions are correctly
chosen. We then turn to {\em on}-shell recursion relations, that deal with the
splitting-up of amplitudes into products of lesser on-shell amplitudes connected by
off-shell propagators. The less-than-$E^0$ behaviour of these amplitudes
then allows us to prove the unitarity of all tree-level amplitudes. For this it will turn out
to be necessary to deform the momenta (and polarisations) of the particles
by extending the four-dimensional phase space of actual physics to
a higher-dimensional one; fortunately, since no spinors are involved the
technicalities of this deformation are fairly straightforward.

\section{The ZH sector in the unitary gauge}
The propagator of the Z and H bosons are given by, respectively,
\bqa
\diag{zprop}{2}{-.35} &=&
{i\over p^2-m^2}\left( -g^{\mu\nu} + {1\over m^2}p^\mu p^\nu\right)\nl
&=& {-i\over p^2-m^2}T^{\mu\nu}(p) + {i\over m^2}L^{\mu\nu}(p)\;\;,\nl
\diag{hprop}{1.6}{-.3} &=& {i\over p^2-M^2}\;\;.
\eqa
Here
\bq
T^{\mu\nu}(p) = g^{\mu\nu} - p^\mu p^\nu/p^2\;\;\;,\;\;\;
L^{\mu\nu}(p) = p^\mu p^\nu/p^2
\eq
are the two purely transverse and longitudinal projection tensors, with $T^2=T$, $L^2=L$ and $TL=0$. The mass of the Z and of the Higgs are denoted by
$m$  and $M$, respectively. 
By either reading them off from the electroweak Lagrangian \cite{bailin},
or by insisting on correct high-energy behaviour of the amplitudes
\footnote{By $\calm(p$Z$,q$H$)$ we denote the tree-level amplitude with $p$
external Z bosons and $q$ external H bosons.} 
$\calm(2$Z,2H), $\calm(2$Z,3H) and $\calm(4$Z,1H) \cite{pvier},
we establish the Feynman rules for the vertices:
\bqa
\diag{zzhvert}{1.5}{-.65} &=& 2ig\,m^2\,g^{\al\be}\;\;\;,\;\;\;
\diag{zzhhvert}{1.5}{-.65}\;\;=\;\;2ig^2\,m^2\,g^{\al\be}\;\;,\nl
\diag{hhhvert}{1.2}{-.65}&=& -3ig\,M^2\;\;\;,\;\;\;
\diag{hhhhvert}{1.2}{-.65}\;\;=\;\;-3ig^2\,M^2\;\;,
\eqa
where
\bq
g^2 = G_F\sqrt{2}\;\;,
\eq
and $G_F$ is the Fermi coupling constant.
For an external Z the longitudinal polarisation vector can be constructed as
\bq
{\eps_L}^\mu = {1\over m}\left( q^\mu - {m^2\over(q\cdot t)}t^\mu\right)\;\;,
\eq
where $t$ is a lightlike vector which we shall choose to be the same for all
Z's in the amplitude. For an arbitrary Minkowski vector $r$, the projection
\bq
r^\mu\;\;\to\;\;r^\mu + {m^2(r\cdot t)-(r\cdot q)(q\cdot t)\over(q\cdot t)^2}t^\mu
 - {(r\cdot t)\over(q\cdot t)}q^\mu
\label{longpolform}
\eq
gives an vector orthogonal to both $q$ and $t$, that can then be normalized to
a transverse polarisation vector $\eps_T$. This is especially useful for Monte
Carlo investigation of the amplitudes.

\section{Recursion relations for off-shell amplitudes}
In order to maximize the power-counting behaviour with $E$ for any
amplitude in the ZH sector, we must choose all external Z's to have
longitudinal polarisation; we must use the $L$ part of the Z propagators;
and we must reduce the number of H propagators to a minimum. This implies
that diagrams with Higgs self-interactions are always of lower order in $E$. 
It is then easily checked that the highest possible $E$ dependence in any
tree amplitude is $E^2$. Since all diagrams in an $n$-point tree amplitude
have the same power $g^{n-2}$ we may put $g=1$ for simplicity. If we
adopt the convention that all external on-shell momenta are counted outgoing, we may
replace the original Feynman rules by the following ones:
\bqa
\diag{zprop1}{1.4}{-.4} &=& \diag{hprop1}{1.4}{-.4}\;\;\;=\;\;\;{i\over p^2}\;\;,\nl
\diag{zzhvert1}{2}{-.5} &=& \diag{zzhhvert1}{2}{-.5}\;\;\;=\;\;\;2i(p\cdot q)\;\;.
\eqa
All external (on-shell) lines carry a trivial factor 1 in this formulation; also
implied is an overall factor $(-)^{n/2}$ in an on-shell amplitude with $n$ external Z
bosons. The ZH model in this limit is a theory with two massless
scalars and a derivative coupling.
We shall compute the off-shell amplitude for a $Z$ or a H
going to $n$ Z's and $k$ H's:
\[
\diag{zampdef}{1.5}{-.35}\;\;\;,\;\;\;\diag{hampdef}{1.5}{-.35}
\]
which we denote by $Z_{n,k}$ and $H_{n,k}$, respectively. 
The amplitude includes the off-shell propagator, and the
momentum $p$ is counted going {\em into\/} the diagrams. The outgoing Higgs momenta are
denoted $h_i$, $i=1,\ldots,k$, and $h=h_1+\cdots+h_k$;
the outgoing Z momenta are denoted by $q_i$, $i=1,\ldots,n$, and
$q=q_1+\cdots+q_n$. 

By explicit calculation for several modest values of $n$ and $k$ we can arrive at 
the following conjecture, which we shall prove:
\bqa
Z_{n,k} &=& \left\{\begin{tabular}{lcr}
$(-)^k\,(n-1+k)!$ &,& $n$ odd \\
0 &,& $n$ even \end{tabular}\right.\;\;,\nl
H_{n,k} &=& \left\{\begin{tabular}{lcr}
1 &,& $n=0,k=1$ \\
$(-)^{k+1}\,(n-2+k)!\,\g_n$ &,& $n\ge2$ even \\
0 &,& $n$ odd\end{tabular}\right.\;\;,\nl
\g_n &=& \left\{\begin{tabular}{lcr}
1 &,& $n=2$\\ $(n-1)!!\,(n-3)!!\,/\,(n-2)!$ &,& $n\ge4$ even\end{tabular}\right.\;\;.
\label{conjecture}
\eqa 
These values can conveniently be gathered into two generating functions:
\bqa
\zeta &=& \zeta(x,y) \;\;=\;\; \suml_{k,n\ge0}\,{x^ny^k\over n!k!}\,Z_{n,k}\;\;=\;\;
{1\over2}\log\left({1+y+x\over1+y-x}\right)\;\;,\nl
\chi &=& \chi(x,y) \;\;=\;\; \suml_{k,n\ge0}\,{x^ny^k\over n!k!}\,H_{n,k}\;\;=\;\;
-1 + \sqrt{(1+y)^2-x^2}\;\;.
\eqa
The SDe for the simplified model giving the $E^2$ terms read
\bqa
\diag{rec3term0}{1.5}{-.3} &=&\diag{rec3terms}{1.0}{.0}\;\;+\;\;\diag{rec3term1}{1.7}{-.85} \;\;+\;\;
\diag{rec3term2}{2.2}{-1.1}\;\;,\nl
\diag{rec4term0}{1.5}{-.4}&=&\diag{rec4terms}{1.}{0}\;\;+\;\;\diag{rec4term1}{1.95}{-1.05} \;\;+\;\;
\diag{rec4term2}{2.9}{-1.1}\;\;,
\label{SDequations}
\eqa
The best calculational strategy is to realise that, in any of our off-shell amplitudes,
the coefficient of $(q_1\cdot q_2)$
must be equal to that of $q^2/2$ in the final expression owing to the symmetry between the
Z's. Similarly the coefficient of $(h_1\cdot h_2)$ gives that of $h^2/2$, and the
coefficient of $(q_1\cdot h_1)$ gives that of $(q\cdot h)$. We therefore only have to
keep track of a few momentum products to be able to reproduce the whole off-shell
amplitude $H_{n,k}$. Similarly, the coefficient of $(p\cdot q_1)$ is that of $(p\cdot q)$, and
the coefficient of $(p\cdot h_1)$ is that of $(p\cdot h)$ in the $Z_{n,k}$.
We can write the off-shell amplitudes as\footnote{The logical step function $\theta(\calp)$ 
is 1 if $\calp$ is true, 0 if $\calp$ is false.}
\bqa
Z_{n,k} &=& \theta\left((n,k)=(1,0)\wph\right) - {2\over p^2}\left((p\cdot q) A^{(1)}_{n,k}
 + (p\cdot h)A^{(2)}_{n,k}\vph\right)\;\;,\nl
H_{n,k} &=&  \theta\left((n,k)=(0,1)\wph\right)- {2\over p^2}\left( {1\over2}q^2 A^{(3)}_{n,k}
 + {1\over2}h^2A^{(4)}_{n,k} + (q.h)A^{(5)}_{n,k}\vph\right) .
\label{SDefore2terms}
\eqa
The several $A$'s are given by
\bqa
 A^{(1)}_{n,k} &=& 
 \suml_{m,\ell} {n-1\choose m-1}{k\choose\ell}\,Z_{m,\ell}\,H_{n-m,k-\ell}\nl &&
+\; \;{1\over2}\suml_{m,t,\ell,r}\,
{n-1\choose m-1\;,\; t}{k\choose \ell\;,\; r}\,
Z_{m,\ell}\,H_{t,r}\,H_{n-m-t,k-\ell-r}\;\;,\nl
A^{(2)}_{n,k} &=& \suml_{m,\ell}{n\choose m}{k-1\choose \ell-1}\,Z_{m,\ell}\,H_{n-m,k-\ell}\nl &&
+ \;\;{1\over2}\suml_{m,\ell,t,r}
{n\choose m\;,\; t}{k-1\choose \ell-1\;,\; r}
\,Z_{m,\ell}\,H_{t,r}\,H_{n-m-t,k-\ell-r}\;\;,\nl
A^{(3)}_{n,k} &=& \suml_{m,\ell} {n-2\choose m-1}{k\choose\ell}\,Z_{m,\ell}\,Z_{n-m,k-\ell}\nl
&&+ \suml_{m,\ell,t,r}{n-2\choose m-1\;,\;t-1}{k\choose\ell\;,\;r}\,Z_{m,\ell}
\,Z_{t,r}\,H_{n-m-t,k-\ell-r}\;\;,\nl
A^{(4)}_{n,k} &=& \suml_{m,\ell} {n\choose m}{k-2\choose\ell-1}\,Z_{m,\ell}\,Z_{n-m,k-\ell}\nl
&&+ \suml_{m,\ell,t,r}{n\choose m\;,\;t}{k-2\choose\ell-1\;,\;r-1}\,Z_{m,\ell}
\,Z_{t,r}\,H_{n-m-t,k-\ell-r}\;\;,\nl
A^{(5)}_{n,k} &=& \suml_{m,\ell} {n-1\choose m-1}{k-1\choose\ell-1}\,Z_{m,\ell}\,Z_{n-m,k-\ell}\nl
&&+ \suml_{m,\ell,t,r}{n-1\choose m-1\;,\;t}{k-1\choose\ell\;,\;r-1}\,Z_{m,\ell}
\,Z_{t,r}\,H_{n-m-t,k-\ell-r}\;\;.
\label{fearsomerecursions}
\eqa
The summations run from 0 to infinity. Throughout this paper we use the convention
\bq
{a\choose b} = 0\;\;\;\mbox{for $b<0$ or $b>a$}
\eq
since $(-n)!$ diverges for natural numbers $n>0$, and the definitions
\bq
{a\choose b\;,\;c} = {a\choose b}{a-b \choose c}\;\;\;,\;\;\;
{a\choose b\;,\;c\;,\;d} = {a\choose b}{a-b\choose c}{a-b-c\choose d}\;\;.
\eq 
To illustrate how the combinatorial factors come about, let us look at the first term
in the recursive expression for $\cala^{(1)}$ in \eqn{fearsomerecursions}.
The coefficient of $(p\cdot q)$ is, as we have argued, that of $(p\cdot q_1)$. Now the
vector $q_1$ can only come from the Z. In the object $Z_{m,\ell}$ we therefore
single out the vector $q_1$, and then there are $m-1$ other external Z momenta left,
to be chosen from $n-1$ available ones; this gives the first binomial. The second
binomial comes from the number of ways to choose $\ell$ H momenta out of $k$.
We stress that we do not assume the momenta $q_j$ to be all equal, as is done in
studies of threshold amplitudes \cite{threshold}; rather, we use the fact that amplitudes must
be symmetric in the $q$'s as well as in the $h$'s.

By computer algebra 
the relations (\ref{SDefore2terms}) can be checked for different values of $(n,k)$;
but it is more profitable to inspect  the generating functions of the $A$'s,
\bq
\cala^{(j)}(x,y) = \suml_{n,k\ge0} {x^ny^k\over n!k!} A^{(j)}_{n,k}\;\;\;,\;\;\;j=1,\ldots,5\;\;.
\eq
The functions $\cala^{(1,2)}$ must be odd in $x$, the functions $\cala^{(3,4,5)}$ even. The SDe
take the following forms:
\bqa
\zeta &=& x -2\cala^{(1)}{(p\cdot q)\over p^2} - 2\cala^{(2)}{(p\cdot h)\over p^2}\;\;,
\label{zetaeqn}\\
\chi &=& y - \cala^{(3)}{q^2\over p^2} - \cala^{(4)}{h^2\over p^2} - 2\cala^{(5)}{(q\cdot h)\over p^2}\;\;.
\label{chieqn}
\eqa
Since we already have our conjecture on the form of $\zeta$ and $\chi$ we only need to
establish the consistency of these equations rather than provide an all-out proof, because 
given the correct starting points the recursion relations lead to unique answers.
We can easily derive the following differential relations:
\bqa
-2\ddx\cala^{(1)} &=& -2\,\left(\ddx\zeta\right)\,\left(\chi + {1\over2}\chi^2\right)\;\;
=\;\;\ddx\zeta - 1 -y\;\;,\nl
-2\ddy\cala^{(2)} &=& -2\,\left(\ddy\zeta\right)\,\left(\chi + {1\over2}\chi^2\right)\;\;
=\;\;\ddy\zeta + x\;\;,\nl
-\ddvv{2}{x}\cala^{(3)} &=& -\left(\ddx\zeta\right)^2\left(1+\chi\vph\right)\;\;
=\;\;\ddvv{2}{x}\chi\;\;,\nl
-\ddvv{2}{y}\cala^{(4)} &=& -\left(\ddy\zeta\right)^2\left(1+\chi\vph\right)\;\;
=\;\;\ddvv{2}{y}\chi\;\;,\nl
-2\ddx\ddy\cala^{(5)} &=& -2\left(\ddx\zeta\right)\left(\ddy\zeta\right)\left(1+\chi\vph\right)\;\;
=\;\;2\ddx\ddy\chi\;\;.
\eqa
Using the even/odd properties of the $\cala$'s we arrive at
\bqa
-2\cala^{(1)} &=& \zeta - x - xy\;\;,\nl
-2\cala^{(2)} &=& \zeta + xy + f_1(x)\;\;,\nl
-\cala^{(3)} &=& \chi  + f_2(y)\;\;,\nl
-\cala^{(4)} &=& \chi + yf_3(x) + f_4(x)\;\;,\nl
-2\cala^{(5)} &=& 2\chi + f_5(x) + f_6(y)\;\;.
\eqa
Here the functions $f_j$ ($j=1,\ldots,6$) are undetermined. Note, however, that the terms
with $f_{1,3,4,5}$ correspond to either $h^\mu=0$ or $h^2=0$, while those with
$f_{2,6}$ correspond to $q^\mu=0$. These terms therefore effectively vanish. 
A sole exception is  the possibility $f_3(x)=$constant. The starting value
$H_{0,1}=1$ tells us to  take $f_3(x)=-1$, so  \eqn{chieqn} is satisfied. The right-hand 
side of \eqn{zetaeqn} reads
\[
\zeta + x\left(1-{(p\cdot q)\over p^2}\right) + xy\left({(p\cdot q)\over p^2}
- {(p\cdot h)\over p^2}\right)\;\;.
\]
For $(n,k)=(1,0)$ we have $p\cdot q = p^2$, and for $(n,k)=(1,1)$ we have
$p\cdot q = p\cdot h = p^2/2$, so that the extra terms also effectively vanish. This
establishes the correctness of the conjecture (\ref{conjecture}). Since the off-shell amplitudes
do not depend on $p^2$ we have now proven that in all on-shell amplitudes the $E^2$
terms vanish.

\section{Mass effects and transversality}
So far  we have taken the Z and H bosons to be massless, which was appropriate for
examining the $E^2$ terms. In the next order, $E^0$, we have to consider the effects of
nonzero masses. These come from different sources. For the Z mass $m$ we
have the effect of $q_j^2=m^2$ for external Z's; the correction term with $t^\mu$ for
longitudinally polarised bosons in \eqn{longpolform}; and the $T$ term in the Z propagator.
Another possible source of $E^1$ or $E^0$ terms is 
the occurrence of one or two transversely polarised Z bosons. For the Higgs mass $M$ we
have the effect of $h_j^2=M^2$ for external Higgses; the correction to the Higgs propagator;
and the so-far neglected H self-interactions. To leading order we may inspect all these 
different effects
separately, while keeping to the $E^2$ approximation in the rest of the amplitude. 

We first deal with the $M^2$ terms. The $M^2$ corrections in the Higgs propagator
can conveniently be included by keeping the propagator massless and including a
two-point vertex:
\bq
\diag{h2vertex}{1.5}{0}\;\;=\;\;-iM^2\;\;.
\eq
Let us now consider an on-shell amplitude with $n$ Z's and $k$ Higgses. The occurrence
of a single 2-,3-, or 4-point vertex gives the $M^2$ contribution to the order we are working in,
denoted by $M(n,k)$.
Keeping track of these vertices and dropping an overall factor $iM^2$, we have
\bqa
M_{n,k} &=& \diag{m2v2}{2}{-.2}\;\; +\;\; \diag{m2v3}{2}{-.9}\;\; +\;\; \diag{m2v4}{1.8}{-.9}\nl
&=& {(-1)\over2}\suml_{m,\ell}{n\choose m}{k\choose\ell}\,H_{m,\ell}H_{n-m,k-\ell}\nl
&+& {(-3)\over3!}\suml_{m,\ell,t,r}
{n\choose m,t}{k\choose \ell,r}\,H_{m,\ell}H_{t,r}H_{n-m-t,k-\ell-r}\nl
&+& {(-3)\over4!}\suml_{m,t,u,\ell,r,s}{n\choose m,t,u}{k\choose\ell,r,s}\,
H_{m,\ell}H_{t,r}H_{u,s}H_{n-m-t-u,k-\ell-r-s}\;\;.\nl
\eqa
The $M^2$ contribution from an external Higgs is correctly subsumed in
the two-point terms. We immediately find the generating function
\bqa
\suml_{n,k}{x^ny^k\over n!k!}M_{n,k} &=& -{1\over2}\chi^2 - {1\over2}\chi^3 - {1\over8}\chi^4\nl
&=& -{1\over8}\chi^2(2+\chi)^2\;\;=\;\; -{1\over8}\left(2y + y^2 - x^2\right)^2\;\;.
\eqa
We find that the $M^2$ terms vanish in all on-shell amplitudes except (correctly) the 
3- and 4-point ones, which after all are not required by unitarity to decrease at high energy.
We can read off the leading $M^2$ terms in these amplitudes:
\bqa
&&\calm(3H) : -3iM^2\;\;,\;\;\calm(2Z,1H) :  +iM^2\;\;,\nl&&\calm(4H) : -3iM^2\;\;,\;\;
\calm(4Z) : -3iM^2\;\;,\;\;\calm(2Z,2H) :  +iM^2\;\;.
\eqa
Explicit calculation confirms these results.
The last of these is the least trivial one: of the 4 diagrams,
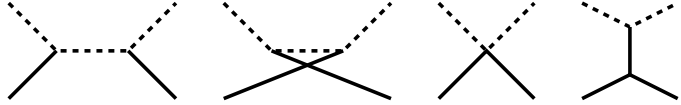
\[\diag{zztohh}{9}{0}\]
each of the first two ones give $-iM^2$,
while the fourth diagram contributes $+3iM^2$. 
We now turn to the $m^2$ terms. To this end another recursion relation is needed, namely that for
off-shell Z's without the propagator factor. These we denote by $K^\mu_{n,k}(q)$, where $\mu$
is the uncontracted Lorentz index of the amplitude. We immediately have, at the $E^2$ level,
\bq
K^\mu_{1,0}(q) = q^\mu\;\;\;,\;\;\;K^\mu_{1,1} = 2iq^\mu\;\;.
\eq
For the other values of $(n,k)$ the recursion relation is diagrammatically also given by the
first line of \eqn{SDequations}. Algebraically we therefore have
\bqa
K^\mu_{n,k}(q) &=& q^\mu\theta(n=1,k=0) + 2iq^\mu\theta(n=1,k=1)\nl
&& +\; 2i\left(
q^\mu A^{(1)}_{n,k} + h^\mu A^{(2)}_{n,k}\wph\right)\theta(n>1\,\mbox{or}\,k>1)\;\;.
\eqa
In terms of generating functions this reads
\bqa
\calk &=& \suml_{n,k}{x^ny^k\over n!k!}K^\mu_{n,k}(q) = xq^\mu + 2ixyq^\mu \nl &&+\;
2iq^\mu\left(\cala^{(1)} - (x,xy)\right) + 2ih^\mu\left(\cala^{(2)} - (x,xy)\right)\;\;,
\eqa
where the notation ``$-(x,xy)$" means that the coefficients of both $x$ and $xy$ are to be
put to zero. We have
\bqa
\cala^{(1)} - (x,xy) &=& \zeta - (x,y)\;\;,\nl
\cala^{(2)} - (x,xy) &=& \zeta + f_7(x) - (x,xy)\;\;,
\eqa
where, as  before, $f_7(x)$ is undetermined but refers to cases with $h^\mu=0$ anyway.
We arrive at 
\bq
\calk = xp^\mu + 2ixyq^\mu + 2i\left(\zeta - (x,xy)\wph\right)p^\mu\;\;.
\label{formofthekmu}
\eq 
With the single exception of the case $n=k=1$, all the amplitudes $K^\mu_{n,k}$ are
seen to be proportional to the momentum $p^\mu$ of the off-shell $Z$ boson. This
has two consequences. In the first place, to this order contracting the amplitude $K^\mu$ with
the Z polarisation vector $\eps_\mu$ gives a vanishing result. The only other
source of $m^2$ terms, the $T$ part of the propagator, is contracted at each end with a $K^\mu$
amplitude. This means that it can only survive if it has $K^\mu_{1,1}$ at both ends; this
implies that the amplitude $\calm(2$Z,2H) is the only one that has $m^2E^0$ terms.\\

From \eqn{formofthekmu} we see that potential $E^1$ terms in amplitudes with one
transversely polarised Z vanish. If {\em two\/} Z bosons are transversely polarised the leading
terms go as $E^0$ by powercounting, and we shall now investigate these by considering
amplitudes having two off-shell legs with 
momenta $q_a$ and $q_b$, and unresolved Lorentz indices $\al$ and $\be$,
respectively, the $n-2$ other Z's being
longitudinally polarized. By Lorentz covariance and power counting, such amplitudes must
be of the form
\[
iQ\,g^{\al\be} + i\sum R_j\,{{a_j}^\al{b_j}^\be\over\De_j}
\]
where $Q$ and the $R_j$ are numbers, $a_j$ and $b_j$ are (combinations of) momenta,
and $\De_j$ is the denominator of some propagator.

The only diagrams that contribute to $g^{\al\be}$ in the amplitudes are those where the
two off-shell Z bosons are connected to the same $ZZH$ or $ZZHH$ vertex. In terms of the
generating functions, we therefore have
\bq
\diag{gterm1}{1.5}{-.5}\;\;+\;\;\diag{gterm2}{1.5}{-.8}\;\; \to\;\;
2ig^{\al\be}\left(\chi + {1\over2}\chi^2\right) = ig^{\al\be}\left(2y + y^2 - x^2\right)\;\;,
\eq
so that $Q=2$ for $\calm(2Z,H)$ and $\calm(2Z,2H)$, $Q=-2$ for $\calm(4Z)$, and
$Q=0$ for all other amplitudes.

The $R_j$ correspond to the residues of the poles of the propagators 
$1/\De_j$. Since neither the $Z_{n,k}$ nor the $H_{n,k}$ have poles, $\De$'s that contain
either none or both the off-shell Z's do not contribute to the $E^0$ terms. Let us consider the
doubly off-shell amplitude 
$\calm(q_a^\al,q_b\be,q_1,q_2,\ldots,q_{n-2},h_1,h_2,\ldots,h_k)$ which has
$n$ Z and $k$ H legs. All Z's except the first two are on-shell and longitudinally polarised;
the Z's with momenta $q_{a,b}$ are off-shell and their Lorentz index is not resolved. The
residue of the pole $(q_a+q_{j_1}+\cdots+q_{j_m}+h_{i_1}+\cdots+h_{i_\ell})^{-2}$
is given, up to a possible sign, by
\bq
R = i\,K^\al_{m+1,\ell}(q_a)\,K^\be_{n-1-m,k-\ell}(q_b)
\eq
for even $m$, and by
\bq
R = i\,K^\al_{m,\ell+1}(q_a)\,k^\be_{n-2-m,k+1-\ell}(q_b)
\eq
for odd $m$. Since $K_{n,k}^\mu(q)$ is proportional to $q^\mu$ except when 
$n=k=1$, each term in the $E^0$ terms of the
amplitude is proportional to either $q_a^\al$ or $q_b^\be$ or both. Upon contraction with
polarisation vectors they therefore cancel. The single exception to this behaviour
is $K_{1,1}(q_a)^\al K_{1,1}^\be(q_b)$, that is the 4-point amplitude 
$\calm(2Z,2H)$ which stands unmasked as the most irregular amplitude of all in the ZH sector.

\section{On-shell recursion relations}
We have now proven the following: all $E^2$ contributions cancel in all $n$-point amplitudes with
$n\ge4$; terms with $E^0M^2$ only survive for $n=3$ or 4;
and terms with $E^0m^2$ only occur in the 2Z,2H amplitude. For $n>4$
all amplitudes  decrease with $E$ at least as fast as $E^{-1}$. To arrive at this
conclusion we have used {\em only\/}  that the external momenta are on-shell, and
momentum conservation. To proceed further
we consider on-shell decomposition relations, in the spirit of \cite{bcfw}.\\

An $n$-point tree amplitude $\calm$ contains $2^{n-1}-n-1$ internal propagators. Let $s$
be a set of $n_s$ of the external momenta (with $2\le n_s\le n-2$), and let us call $p_s$ the total
momentum of the set $s$. The corresponding propagator has denominator
\bq
\Delta_s = {p_s}^2 - m^2\;\;,
\eq
where $m$ is the Z or Higgs mass, as the case may be.
We shall describe a deformation of the amplitude into a phase space of dimension 7 (a higher
number is in principle also possible).
The metric has signature
$(+,-,-,-,+,-,-)$. The external momenta $k_j$ ($j=1,\ldots,n)$ have of course no
components in the extra dimension, nor does the auxiliary vector $t$:
\bqa
&&k^\mu = (k^0,k^1,k^2,k^3)\;\;\;\to\;\;\;(k^0,k^1,k^2,k^3,0,0,0)\;\;,\nl
&&t^\mu = (t^0,t^1,t^2,t^3)\;\;\;\to\;\;\;(t^0,t^1,t^2,t^3,0,0,0)\;\;.
\eqa
Now, we choose an arbitrary set of vectors $\eta_j$ ($j=1,\ldots,n)$ having
components {\em only\/} in the extra dimensions:
\bq
\eta^\mu = (0,0,0,0,\eta^5,\eta^6,\eta^7)\;\;,
\eq
with the constraints
\bq
{\eta_j}^2 = 0\;\;\;,\;\;\;\suml_j{\eta_j}^\mu = 0\;\;.
\eq
It is this requirement that necessitates the extra dimensions to number at least 3;
but then we can always construct any number of such vectors.
We define the following deformation, depending on a complex parameter $z$:
\bq
{k_j}^\mu \to {\ktil_j}^\mu\;\;=\;\;{k_j}^\mu + z^{1/2}{\eta_j}^\mu
\eq
The longitudinal polarization vector of an external Z now automatically gets the
deformation
\bq
{\eps_j}^\mu \to {\etil_j}^\mu = {\eps_j}^\mu + {z^{1/2}\over m}{\eta_j}^\mu\;\;.
\eq
Note, however, that there are not 2 but 5 transverse polarisation vectors, with
components in all dimensions.
Just like the original $k_j$, the deformed $\ktil_j$ are on-shell and the total momentum is
conserved.

The deformed amplitude, $\calm(z)$, is the original one, $\calm(0)$, with the momenta and
polarisations replaced by their deformed versions. It has denominators
\bq
\tilde{\Delta}_s = {\tilde{p}_s}^2 - m^2 = \Delta_s + z{\eta_s}^2\;\;,
\eq
which vanish at the $z$ value
\bq
z_s = - \Delta_s/{\eta_s}^2\;\;.
\eq
The residue at this pole is denoted $R(z_s)$.
It is easily seen that $\calm(z)$ is a rational function of $z$. Let us consider the limit $z\to\infty$.
In this limit, the $\eta$ dominate the external momenta. Since the $E^2$ terms vanish, as do the
$E^{1,0}$ terms for $n>4$,  
\bq
\calm_\infty \equiv \lim_{z\to\infty}\calm(z) = \left\{
\begin{tabular}{ccl} \mbox{constant} &,& $n=4$\\
                               0 &,& $n>4$
\end{tabular}\right.
\eq          
We are therefore allowed the following contour integral manipulations:
\bqa
\calm(0) &=& {1\over2\pi i} \oint\limits_{z\sim 0} dz\; {\calm(z)\over z}\nl
&=& \calm_\infty - \suml_s{1\over2\pi i}
\oint\limits_{z\sim z_s} dz\;{R(z_s)\over z\,\tilde{\Delta}_s}\nl
&=& \calm_\infty - \suml_s{1\over2\pi i}
\oint\limits_{z\sim z_s} dz\;{R(z_s)\over z(z-z_s){\eta_s}^2}\nl
&=& \calm_\infty + \suml_s {R(z_s)\over\Delta_s}\;\;.
\eqa
Since at the pole $z=z_s$ the deformed momentum $\tilde{p}_s$ is exactly on-shell, and
since for Z propagators in the unitary gauge
\bq
\left\lfloor-g^{\mu\nu} + {1\over m^2}\tilde{p}_s^\mu\tilde{p}_s^\nu\right\rfloor_{z=z_s} = 
\suml_{\la=1}^6 \etil_s(\la)^\mu \etil_s(\la)^\nu\;\;,
\eq
where $\la$ denotes an enumeration of the (at least) six physical, normalised polarisation vectors,
we see that $R(z_s)$ is precisely (a spin sum of) the product of two on-shell amplitudes:
\bq
R(z_s) = A_{n_s+1}B_{n-n_s+1}\;\;,
\eq
where we have indicated the number of external legs in the factor amplitudes, which is always
at most $n-1$. This allows us induction in $n$: if both the on-shell amplitudes $A$ and $B$
respect unitarity in the sense that they have the correct behaviour with $E$, then
\bq
{R(z_s)\over\Delta_s} \sim {1\over E^2}\,E^{4-(n_s+1)}\,E^{4-(n-n_s+1)}
= E^{4-n}\;\;.
\eq
Thus we have established that in the HZ model all on-shell tree amplitudes obey
partial-wave unitarity.

A final remark is in order here. We want to stress that in this paper we do not aim at
{\em computing\/} the amplitudes, but rather want to study their high-$E$ behaviour.
If we had opted for a two-line deformation, we would in the limit $|z|\to\infty$ have the situation
of two high-energy particles moving in the background of lower-energy ones. The high-$E$
limit would then be a situation  like $m,M \ll q_{2,\ldots,n},h_{2,\ldots,n} \ll zq_1,zh_1$, 
a problem in which {\em two\/} large ratios of scales occur. By using an all-line
deformation we circumvent this artificial problem because the high-$E$ and high-$z$
limits actually conicide. Other all-line deformations have been used before
\cite{deformations}, where the fact that Weyl spinors are used more or less enforces the
restriction to four dimensions. Since we only consider vectors, the extension to higher
dimensions is unproblematic. On the other hand, higher dimensions imply extra transverse
polarisations, which are of course absent in a four-dimensional deformation. For internal
lines, the only r\^ole of the extra polarisations is to ensure that the Z propagators
remain in the unitary gauge; while for the external lines, only the `original' three
polarizations are present, albeit deformed\footnote{We thank the referee for drawing our
attention to this point.}.

\section{Conclusions}
In this paper we prove the tree-level unitarity of all amplitudes in the Abelian Higgs model.
This is not a new result: rather, it is the method of arriving at it that is
of interest here, and we recapitulate the novelties involved.
\begin{itemize}
\item We have used only physical fields.
The unitary gauge is widely considered inappropriate for studying
unitarity (and renormalizability) because of its high-energy behaviour, but here we have shown
that it actually forms the cornerstone of any treatment that aims at using {\em physical\/}
degrees of freedom only: it provides the effective Feynman rules that led us to $Z_{n,k}$ and
$H_{n,k}$. 
\item The Schwinger-Dyson equations of the theory are seen to lead to surprsingly
simple forms for the off-shell amplitudes ({\it cf\/} \eqn{conjecture}), which
have to our knowledge not been obtained before.
\item We have deformed the amplitudes by extending the dimensionality of phase space
and deforming all lines simultaneously. We deem this all-line deformation necessary since
we are dealing with a massive theory rather than unbroken YM-like theories in which the
problem of relative scales does not enter. 
\end{itemize}


\begin{thebibliography}{999}
\bibitem{Abelianhiggsmodel}P. W. Anderson, Phys. Rev. {\bf130} (1962) 439.

\bibitem{tHV} G. 't Hooft, M. Veltman, Nucl. Phys. {\bf B 44} (1972) 189.

\bibitem{brs} C.Becchi, A.Rouet, R. Stora, Phys.Lett 52{\bf{B}} (1974) 344.

\bibitem{HelV}
H.G.J. Veltman, Phys.Rev. D41 (1990) 2294. 

\bibitem{knetter}
C. Grosse-Knetter and R. K\"ogerler, Phys.Rev. {\bf D}48 (1993) 2865. 

\bibitem{elvangreview}
H. Elvang and Y.-T. Huang, arXiv:1308.1697.

\bibitem{weinzierl}
S. Weinzierl, Phys.Rept. 676 (2017) 1.

\bibitem{zuber}
C. Itzykson and J.-B. Zuber, {\it Quantum Field Theory} (1980).

\bibitem{peskin}
M. E.  Peskin and D. V. Schroeder, {\it An Introduction to Quantum Field Theory} (1995).
 
\bibitem{bailin} A most complete set of Feynman rules is given in {\it e.g.\/} D. Bailin and A. Love, {\it Introduction to Gauge Field Theory\/}
(IOP and Adam Hilger, 1986).

\bibitem{pvier} R.Kleiss, `P4' lecture notes `Paths, Pictures, Particles, Processes',
Radboud University Nijmegen 2017. This approach follows that of
J. M. Cornwall, D. N. Levin and G. Tiktopoulos, Phys.Rev.Lett. 30 (1973) 1268 (Erratum: Phys.Rev.Lett. 31 (1973) 572), and
Phys.Rev. {\bf D}10 (1974) 1145 (Erratum: Phys.Rev. {\bf D}11 (1975) 972).

\bibitem{threshold}
For example, in
E.N. Argyres, C. G. Papadopoulos and R. Kleiss Phys.Lett. {\bf B}302 (1993) 70;
Phys.Lett. {\bf B}319 (1993) 544;
L. S. Brown, Phys. Rev. {\bf D}46 (1992) 4125. For a recent application, see for instance
V. V. Khoze and M. Spannowsky, arXiv:1704.03447.

\bibitem{bcfw}
R. Britto, F. Cachazo, B. Feng, E. Witten,  Phys.Rev.Lett. {\bf 94}(2005) 181602. 

\bibitem{deformations}
N. Arkani-Hamed, J. Kaplan, JHEP 0804 (2008) 76;
H. Elvang, D. Z. Freedman and M. Kiermaier, JHEP 0906 (2009) 68;
T. Cohen, H. Elvang and M. Kiermaier, JHEP 1104 (2011) 53.

\end{thebibliography}
\end{document}